\documentclass[letterpaper,  
              ]{jacow}
\makeatletter%
	\ifboolexpr{bool{xetex}}
	 {\renewcommand{\Gin@extensions}{.pdf,%
	                    .png,.jpg,.bmp,.pict,.tif,.psd,.mac,.sga,.tga,.gif,%
	                    .eps,.ps,%
	                    }}{}
\makeatother

\ifboolexpr{bool{xetex} or bool{luatex}} 
 {}                                      
 {\usepackage[utf8]{inputenc}}           

\usepackage[english]{babel}			 

\usepackage[final]{pdfpages}
\usepackage{multirow}
\usepackage{ragged2e}
\usepackage{amsmath}
\ifboolexpr{bool{jacowbiblatex}}%
 {%
  \addbibresource{jacow-test.bib}
  \addbibresource{biblatex-examples.bib}
 }{}
\usepackage{tikz}
\usepackage{tikz-dsp}
\usepackage{relsize}
\usepackage{amsmath}
\usetikzlibrary{shapes,arrows,decorations.markings,chains}
\tikzstyle{vecArrow} = [thick, decoration={markings,mark=at position
	1 with {\arrow[semithick]{open triangle 60}}},
double distance=1.0pt, shorten >= 5.5pt,
preaction = {decorate},
postaction = {draw,line width=1.0pt, white,shorten >= 2.5pt}]
\tikzstyle{innerWhite} = [semithick, white,line width=1.0pt, shorten >= 2.5pt]

\DeclareMathAlphabet{\mathpzc}{OT1}{pzc}{m}{it}

\tikzstyle{block} = [draw, rectangle, 
minimum height=0.5mm, minimum width=0.5mm]
\tikzstyle{sum} = [draw, circle, node distance=1mm]
\tikzstyle{input} = [coordinate]
\tikzstyle{output} = [coordinate]
\tikzstyle{pinstyle} = [pin edge={to-,thin,black}]

\begin{document}

\title{Automatic phase calibration for RF cavities using beam-loading signals}

\author{J. P. Edelen\thanks{Now at RadiaSoft LLC, Boulder, CO, jedelen@radiasoft.org}\thanks{The authors of this work grant the arXiv.org and LLRF Workshop's International Organizing Committee a non-exclusive and irrevocable license to distribute the article, and certify that they have the right to grant this license.}, B. E. Chase, Fermilab, Batavia, IL }
	
\maketitle

\begin{abstract}
Precise calibration of the cavity phase signals is necessary for the operation of any particle accelerator. For many systems this requires human in the loop adjustments based on measurements of the beam parameters downstream. Some recent work has developed a scheme for the calibration of the cavity phase using beam measurements and beam-loading however this scheme is still a multi-step process that requires heavy automation or human in the loop. In this paper we analyze a new scheme that uses only RF signals reacting to beam-loading to calculate the phase of the beam relative to the cavity. This technique could be used in slow control loops to provide real-time adjustment of the cavity phase calibration without human intervention thereby increasing the stability and reliability of the accelerator. 
\end{abstract}

\section{introduction}
For modern accelerators precise control of the the amplitude and phase of RF cavities is necessary in order to meet the machine performance requirements. The state of the art in LLRF control is capable of achieving local stability in the amplitude and phase of RF cavities to better than 10$^{-4}$
\cite{ref1}, however this does not guarantee that the cavity amplitude and phase are properly calibrated to the energy gain and synchronous phase of the beam respectively. Additionally sources of drift such as temperature, humidity, and up-stream beam motion, can shift the calibration of the cavity over time and lead to unwanted beam parameters in the accelerator. While there are many calibration schemes in use
\cite{ref2,ref3,ref4,ref5,ref6,ref7,ref8}, such as time-of-flight, spectrometer, or analysis of the beam-loading signals in the RF cavity, all of these methods currently require a human in the loop and machine time in order to update these calibrations and compensate for drift. Therefore the development of an automatic calibration scheme that can be done on the fly would be highly beneficial to machine operation. 

In this paper we will describe our method for measuring the beam phase relative to the RF cavity and show proof-of-priniciple results both with an ideal beam-like disturbance and with a real beam disturbance. We show that for an ideal beam we can calculate the phase of the disturbance relative to the RF to within 0.5 degrees. However, when applied with real beam we have large systematic errors that prohibit the use of this technique on these particular cavities. The source of the error is still unknown but it could be attributed to low-beta effects, parasitic modes, or other non-ideal properties of this particular beam-cavity interaction.

\section{Theory} 
The analysis in this section will follow a block diagram model of the RF control system, Figure 1, which has been shown to compare well with measurements of the RF system. For this model we consider a system with in-phase and quadrature control loops that are independent. Because these two loops are identical we will only analyze one in detail. When using this model for computing the beam-loading phase it is important to understand our underlying assumptions:
\begin{itemize} 
\item The amplifiers are in the linear regime 
\item The loop phase and gain are calibrated 
\item Disturbances other than beam-loading are relatively small or slow compared with beam-loading. 
\end{itemize} 
During this section we will translate between frequency and time domain, functions $f(t)$ in the frequency domain will be noted by $\hat{f}(s)$. 

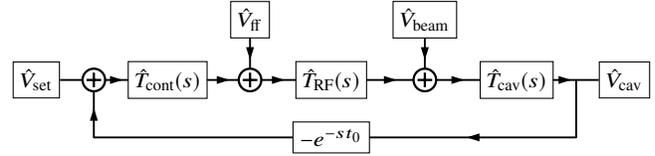
\begin{figure} [h]
	\centering
	\begin{tikzpicture}[scale=0.09]
	\smaller
	\matrix (m1) [row sep=3mm, column sep=3mm]
	{
		\node[coordinate]                  (m200) {};          
		\node[coordinate]                  (m202) {};          &
		\node[coordinate]                    (m204) {};          &
		\node[coordinate]                  (m205) {};          &
		\node[block]                  (m208){$\hat{V}_\mathrm{ff}$};          &
		\node[coordinate] 		(m2087){}; &
		\node[block]                    (m207) {$\hat{V}_\mathrm{beam}$};          &
		\node[coordinate]                    (m209) {};          &		
		\node[coordinate]                    (m210) {};          &\\
		\node[block]                  (m000) {$\hat{V}_\mathrm{set}$};          &
		\node[dspadder,minimum size=3mm]                  (m002) {};          &
		\node[block]                    (m004) {$\hat{T}_\mathrm{cont}(s)$};          &
		\node[dspadder,minimum size=3mm]                  (m005) {};          &
		\node [block] (m006) {$\hat{T}_\mathrm{RF}(s)$};&
		\node[dspadder,minimum size=3mm]                  (m007) {};          &
		\node[block]                    (m008) {$\hat{T}_\mathrm{cav}(s)$};          &
		\node[coordinate]                  (m009) {};          &
		\node[block]                  (m010) {$\hat{V}_\mathrm{cav}$};          & \\
		\node[coordinate]                  (m100) {};          &
		\node[coordinate]                  (m102) {};          &
		\node[coordinate]                    (m104) {};          &
		\node[coordinate]		(m105) {}; &
		\node[block]                    (m106) {$-e^{-st_0}$};          &
		\node[coordinate]                    (m107) {};          &
		\node[coordinate]                  (m108){};          &
		\node[coordinate]                    (m109) {};          &		
		\node[coordinate]                    (m110) {};         & \\
	};
	
	\begin{scope}[start chain]
	\chainin (m000); 
	\chainin (m002) [join=by dspline];
	\chainin (m004) [join=by dspflow];
	\chainin (m005) [join=by dspflow];
	\chainin (m006) [join=by dspflow];
	\chainin (m007) [join=by dspflow];
	\chainin (m008) [join=by dspflow];
	\chainin (m009) [join=by dspflow];
	\chainin (m109) [join=by dspline];
	\chainin (m106)[join=by dspflow];
	\chainin (m102)[join = by dspline];
	\chainin(m002)[join=by dspflow];
	\end{scope}
	
	\begin{scope}[start chain]
	\chainin (m208);
	\chainin (m005) [join = by dspflow];
	\end{scope} 
	

	\begin{scope}[start chain]
	\chainin (m207);
	\chainin (m007) [join=by dspflow];
	\end{scope} 
	
	\begin{scope}[start chain]
	\chainin (m009);
	\chainin (m010) [join=by dspline];
	\end{scope} 
	\end{tikzpicture}
	\caption{Base-band model of LLRF the controller} 
\end{figure}
Using block diagram analysis and the assumptions noted above, we derive the relationship between the cavity voltage $V_\mathrm{cav}$ and the various inputs to the RF control system, Equation 1. 
\begin{multline}
	\hat{V}_\mathrm{cav} = \left(\hat{V}_\mathrm{set}\hat{T}_\mathrm{RF}(s)\hat{T}_\mathrm{cont}(s) + \hat{V}_\mathrm{ff} \hat{T}_\mathrm{RF}(s) + \hat{V}_\mathrm{beam}\right) ... \\
	  \left(\hat{T}_\mathrm{cav}(s) \over 1+\hat{T}_\mathrm{cav}\hat{T}_\mathrm{RF}(s)\hat{T}_\mathrm{cont}(s)e^{-st_0} \right)  
\end{multline}

If we assume that $V_\mathrm{set}(t)$ and $V_\mathrm{ff}(t)$ are not changing during the beam-loading transient and that the cavity voltage is at steady state before the beam arrived, then we can simply subtract off their associated contributions to the cavity field at a given time during the beam-loading transient, for example by sampling the cavity field before the beam arrives $V(0)$. Next we will use the result from Equation 1 to calculate the in-phase and quadrature components of the beam-induced cavity signal. At this point it is convenient to express the transfer function seen by the beam as an arbitrary function $\hat{F}(s)$. We will show later that the details of this transfer function are irrelevant to calculating the beam-phase relative to the RF cavity. This gives the relationship between the cavity voltage and the beam disturbance as $\hat{V}(s) = V(0)/s + \hat{V}_\mathrm{beam}(s) \hat{F}(s)$. Next we convert to time domain to simplify the analysis of the cavity field signals inside the LLRF system, Equation 2. 

\begin{equation}
V_\mathrm{cav}(t) = V_\mathrm{cav}(0) + \int_{-\infty}^{\infty} V_\mathrm{beam}(t)F(t-\tau)d\tau
\end{equation}

Next we assume that the variation in the beam voltage along the beam pulse is small, or that the beam pulse can be approximated as a top-hat disturbance, $V_\mathrm{beam}(t) = V_0 (u(t) - u(t - L))$, where $L$ is the length of the beam pulse, and $V_0$ is the magnitude of the beam voltage. Substituting this representation of the beam voltage into Equation 2 gives $V_\mathrm{cav}(t) = V_\mathrm{cav}(0) + V_0 \int_{-\infty}^{\infty}(u(t) - u(t - L))F(t-\tau)d\tau$. If $V_\mathrm{beam}$ has in-phase and quadrature components then the beam phase is simply, $\arctan(V_0^Q / V_0^I)$. Now that $V_0$ is outside the integral we can solve for it directly in terms of the system transfer function and the cavity voltage, 
\begin{equation} 
V^{\smaller{I,Q}}_0 = { V^{\smaller{I,Q}}_\mathrm{cav}(t) - V^{\smaller{I,Q}}_\mathrm{cav}(0) \over \int_{-\infty}^{\infty}(u(t) - u(t - L))F(t-\tau)d\tau }.
\end{equation}
When we compute the cavity phase from the definition given above, the denominator of Equation 3 will fall out and the beam phase is represented completely by the measured cavity signals, 
\begin{equation} 
\phi_\mathrm{beam} = \tan^{-1}\left( V^{\smaller{Q}}_\mathrm{cav}(t) - V^{\smaller{Q}}_\mathrm{cav}(0) \over V^{\smaller{I}}_\mathrm{cav}(t) - V^{\smaller{I}}_\mathrm{cav}(0) \right).
\end{equation}
To improve signal-to-noise it may be desirable to Integrate the top and bottom of Equation 4 from the start of beam to some point during the beam pulse.

\section{Overview of test cavity} 
This algorithm was tested on room temperature bunching cavities installed at the PIP-II Injector test, Figure 2. These cavities are quarter wave resonators that operate at 162.5 MHz. These cavities have a loaded quality factor of \~5000, an $r/Q$ of \~600, an operating voltage between 50 kV and 100 kV and a beam induced voltage of approximately 15 kV. The LLRF system for these cavities is a base-band I/Q controller that can operate in both pulsed mode and CW mode. In pulsed mode there is data acquisition at a rate of 1 MHz.
\begin{figure} [h]
	\centering
	\includegraphics[width=0.5\textwidth]{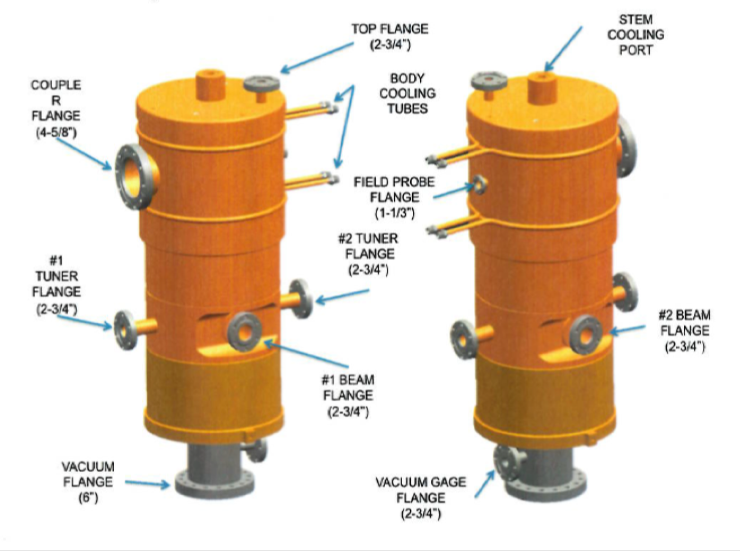}
	\caption{Schematic of the quarter wave bunching cavity used for testing}
\end{figure}
Testing of this scheme was performed in two stages, the first was using a beam like disturbance driven from the LLRF system and the second was using real beam loading signals. In the next two sections we discuss the setup and results from these two tests. 

\section{Ideal beam-like disturbance}

The first test is to use an ideal beam-like disturbance that is driven from the LLRF system. This is accomplished by adding a top-hat profile to the feed-forward table in the center of pulse. Then using Equation 4 we calculated the phase of the disturbance relative to the field in the cavity. This calculated phase was then compared to the set point phase for the feed-forward disturbance in the LLRF system. Figure 3 shows the difference between the calculated phase from the cavity signals and the set point in the RF system. 
\begin{figure} [h] 
	\centering
	\includegraphics[width=0.5\textwidth]{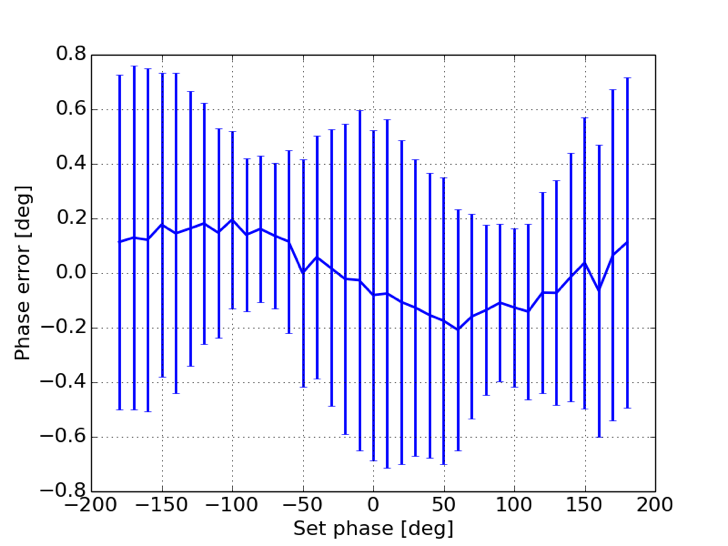}
	\caption{Difference between calculated phase of LLRF disturbance and the set-point phase as a function of set-point phase }
\end{figure}
Here we see fairly good agreement between the calculated disturbance phase and the set point phase, but there is a significant amount of noise in the measurement that is correlated ith the set-point phase. This is believed to be caused by cross-talk in the digital processing board. However on average the error is less than 0.2$^\circ$ peak to peak, which is an encouraging result for this scheme. 

\section{Beam loading tests} 

Next we attempted to apply this technique to real beam loading signals. For this study we scanned the reference phase of the RF system, effectively scanning the phase of the beam relative to the cavity, and then used the beam loading signals to calculate the beam phase relative to the RF system. Because we are only looking at relative phase we subtracted out the linear relationship between the calculated beam phase and the reference phase to understand how well this technique will work when applied to real beam loading signals. 
\begin{figure} [h]
	\centering
	\includegraphics[width=0.5\textwidth]{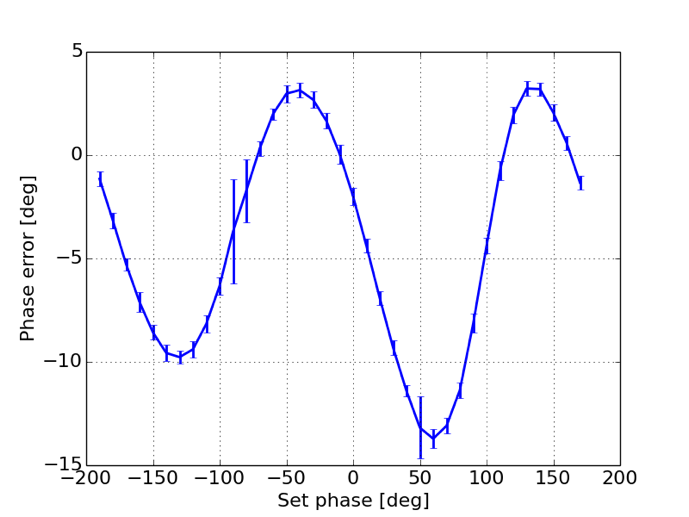}
	\caption{Residual phase error from calculated beam phase using equation 4 as a function of reference phase}
\end{figure}
Here we can see that while we have relatively low statistical errors in the beam phase calculation, there are significant correlated errors as we scan the reference phase. After some investigation the source of the correlated errors is still under discussion. Some possibilities include parasitic modes being excited by the beam, non-relativistic effects, or other non-ideal effects in our system. In order to fully understand these effects, more work is needed. Additionally, implicit in our calculations is that the beam-loading disturbance is sinusoidal with phase. It is possible that this is not the case for different cavity types.

\section{Conclusions} 
In conclusion, we have proposed a method that can in principle determine the phase of the beam relative to the RF system which gives the true synchronous phase of the beam to within $0.5^\circ$. This method could be used in an online updating scheme that regularly adjusts the calibration of the cavity to keep the cavity phase calibration relative to the synchronous phase to within $0.5^\circ$. While testing with beam showed results that are not ideal, we are encouraged that this technique applied to other cavities could yield fruitful results.

\section{Acknowledgments} 
Operated by Fermi Research Alliance, LLC under Contract No. De-AC02-07CH11359 with the United States Department of Energy


\begin{thebibliography}{1}
\bibitem{ref1} J. Branlard, \emph{Low Level RF for SRF Accelerators} Proceedings of LINAC 2014, Geneva Switzerland

\bibitem{ref2} M. G. Tienfenback and K. Brown, \emph{Beam Based Phase Monitoring and Gradient Calibration of Jefferson Laboratory RF Systems} Proceedings of PAC 1997

\bibitem{ref3}  S. L. Kramer and J-M. Wang, \emph{Beam Indued RF Cavity Transient Voltage}, BNL-64886-99/10-Rev, Oct 1998

\bibitem{ref4}  S Ryden, \emph{Beam Loading Studies or Cavity Phase and Amplitude Setting}, M.S. Thesis, Department of Electrical and Information Technology, Lund University

\bibitem{ref5} M. Omet et al, \emph{Development and Application of a Frequency scan-based and a Beam-Based Calibration Method for the LLRF Systems at KEK STF} Proceedings of Particle Accelerator Society of Japan
9th Annual meeting

\bibitem{ref6}  R. Zeng, and O. Troeng, \emph{Transient Beam Loading Based Calibration for Cavity Phase and Amplitude Setting} Proceedings of HB 2016, Malmo Sweden

\bibitem{ref7}  P. Pawlik et al, \emph{New Method for beam induced transient measurement}, Measurement Science and Technology, 18 (2007) 2348-2355

\bibitem{ref8} 
  S.~N.~Simrock and T.~Schilcher,
  \emph{Transient beam loading based calibration of the vector-sum for the
  TESLA test facility} Proceedings of EPAC 1996, Sitges, Spain.

\end{thebibliography}
\end{document}